\documentclass[10pt]{article}

\usepackage{amssymb}
\usepackage{amsfonts}
\usepackage{amsmath}
\usepackage{natbib}
\usepackage{setspace}
\usepackage{booktabs}
\usepackage{authblk}
\usepackage{tikz}
\usepackage{graphicx}
\usepackage{bbm}
\usepackage{xcolor}
\usepackage{apalike}

\graphicspath{ {Figures/} }

\begin{document}
\title{Bayesian Inference for Population Attributable Measures from Under-identified Models}
\author[1*]{Sarah Pirikahu}
\author[2]{Geoffrey Jones}
\author[3]{Martin L. Hazelton}
\affil[1]{School of Population and Global Health,  University of Western Australia,  Perth,  Australia. *sarah.pirikahu@uwa.edu.au}
\affil[2]{School of Fundamental Sciences - Statistics,  Massey University,  Palmerston North, New Zealand}
\affil[3]{Department of Mathematics \& Statistics,  University of Otago,  Dunedin,  New Zealand}
\date{}
\setcounter{Maxaffil}{0}
\renewcommand\Affilfont{\small}

\maketitle

\begin{abstract}
Population attributable risk (PAR) is used in epidemiology to predict the impact of removing a risk factor from the population. Until recently, no standard approach for calculating confidence intervals or the variance for PAR was available in the literature.  \cite{Pirikahu2016} outlined a fully Bayesian approach to provide credible intervals for the PAR from a cross-sectional study, where the data was presented in the form of a  $2 \times 2$ table. However, extensions to cater for other frequently used study designs were not provided. In this paper we provide methodology to calculate credible intervals for the PAR for case-control and cohort studies. Additionally, we extend the cross-sectional example to allow for the incorporation of uncertainty that arises when an imperfect diagnostic test is used. In all these situations the model becomes over-parameterised, or non-identifiable, which can result in standard ``off-the-shelf'' Markov chain Monte Carlo updaters taking a long time to converge or even failing altogether. We adapt an importance sampling methodology to overcome this problem, and propose some novel MCMC samplers that take into consideration the shape of the posterior ridge to aid in the convergence of the Markov chain.
\end{abstract}

\section{Introduction}

The population attributable risk (PAR) is used in epidemiology to predict the impact of a proposed intervention (the removal of a risk factor) on the disease burden of a population. The PAR can be defined as
\begin{equation*}\label{eqn:PAR1}
PAR = P(D^+) - P(D^+|E^-), 
\end{equation*}
where $D^+$ denotes disease presence and $D^-$ disease absence, and similarly $E^+$ and $E^-$ denote exposure status to the risk factor \citep{MacMahon1960}. Alternatively, the PAR can be expressed in terms of the population parameters $p = P(D^+|E^+)$, $q=P(D^+|E^-)$ and $e=P(E^+)$ as follows
\begin{equation}\label{eqn:PAR2}
PAR = e(p - q).
\end{equation}
A similar measure, the population attributable fraction (PAF) defined simply as the PAR divided by $P(D+)$, was proposed by \cite{Levin1953} and several equivalent mathematical definitions of the PAF can be seen thoughout the literature \citep{Rockhill1998}. The similarity in nomenclature of these attributable measures, and often lack of clear mathematical definition and assumptions being made, has resulted in confusion \citep{Rockhill1998, Greenland1988,  Uter2001}. In particular the structured literature search performed by \cite{Uter2001} on 334 papers between 1966 and 1996 showed that 65\% of authors provided no exact definition for their attributable measures used and only 19\% provided confidence intervals.

The confusion in the literature surrounding population attributable measures has probably contributed to the lack of a standard methodology for estimating the uncertainty for the PAR in particular. \cite{Newson2013} provided a module in the statistical program STATA to estimate the PAR and its corresponding confidence interval, but did not take into consideration the uncertainty in the prevalence of the risk factor or clearly address the underlying study design.  Alternative Frequentist and Bayesian approaches for estimating confidence intervals for the PAR when data is in the form of a $2 \times 2$ table from cross-sectional studies is provided by \cite{Pirikahu2016}. However, extensions to cater for other frequently used study designs, such as case-control and cohort studies, were not provided. 

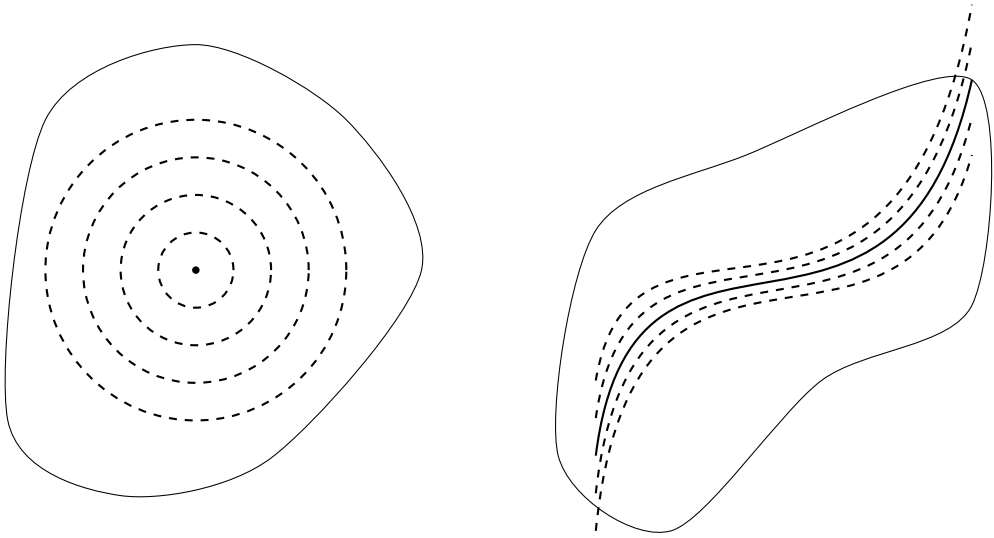
\begin{figure}
	\centering
	\begin{minipage}{.45\textwidth}
		\begin{tikzpicture}
		\draw [black] plot [smooth cycle] coordinates {(-0.5,0) (1,-1) (3,-0.5) (5,2) (4,4) (2,5) (0,4)};
		\draw[black,thick,dashed] (2,2) circle (2cm);
		\draw[black,thick,dashed] (2,2) circle (1.5cm);
		\draw[black,thick,dashed] (2,2) circle (1cm);
		\draw[black,thick,dashed] (2,2) circle (0.5cm);
		\node at (2,2) [circle,fill,inner sep=1pt]{};
		\end{tikzpicture}
	\end{minipage}
	\begin{minipage}{.45\textwidth}
		\begin{tikzpicture}
		\draw [black] plot [smooth cycle] coordinates {(-0.5,0) (1,-1) (3,1) (5,2) (5,5) (2,4) (0,3)};
		\draw[black,thick,dashed] (0,1) .. controls (0.5,4) and (4,0.5) .. (5,6);
		\draw[black,thick,dashed] (0,0.5) .. controls (0.5,4) and (4,0.5) .. (5,5.5);
		\draw[black,thick] (0,0) .. controls (0.5,4) and (4,0.5) .. (5,5);
		\draw[black,thick,dashed] (0,-0.5) .. controls (0.5,4) and (4,0.5) .. (5,4.5);
		\draw[black,thick,dashed] (0,-1) .. controls (0.5,4) and (4,0.5) .. (5,4);
		\end{tikzpicture}
	\end{minipage}
	\caption{Illustration of identifiability issues where the outer most solid lines represent the parameter space. \textit{Left:} Identifiable model where the central dot represents the maximum likelihood estimate and dashed lines the likelihood contours. \textit{Right:} Non-identifiable model where the solid line represents the set of values which have the same maximum likelihood and the dashed lines the likelihood contours.}
	\label{fig:Contours}
\end{figure}

In this paper we provide a fully Bayesian methodology to calculate credible intervals for the PAR for case-control and cohort studies. Our methodology allows for experts to incorporate prior knowledge on either the prevalence of disease or the probability of exposure to the risk factor being considered for removal. Additionally, we extend the cross-sectional study example previously explored by \cite{Pirikahu2016} to allow for the incorporation of uncertainty that arises when an imperfect diagnostic test is used. In all these situations the model becomes over-parameterised, or non-identifiable, meaning that there exists multiple values for the parameter vector of interest that produce the same probability distribution for the observed values,  creating a ``ridge'' in the parameter space (Figure~\ref{fig:Contours}, right). For identifiable models, as the sample size increases, which can be represented by the dashed lines in Figure~\ref{fig:Contours}, the likelihood contours will shrink towards a point, i.e. the maximum likelihood estimate (Figure~\ref{fig:Contours} , left). When the model is non-identifiable standard Frequentist methods for parameter estimation are not a viable option. Under a Bayesian framework, the addition of prior information in the form of a proper prior leads to a proper posterior distribution, so inference is possible. Obtaining the posterior distribution however, can be problematic as standard Markov chain Monte Carlo (MCMC) algorithms, such as Gibbs and Metropolis-Hastings, can take a long time to converge or even fail altogether \citep{Gustafson2015}. The problem gets worse as the data size gets bigger. To estimate the PAR from a cross-sectional study which relies on an imperfect diagnostic test of disease or exposure, we adapt a general importance-sampling approach developed by \cite{Gustafson2015} for non-identifiable models. We also propose some novel MCMC samplers for use if finding a ``transparent re-parameterisation'' required for Gustafson's importance sampling is problematic; these samplers take into consideration the shape of a posterior ridge to aid in the convergence of the Markov chain.  Each of the methods for the cross-sectional study incorporating diagnostic testing are compared in terms of their efficiency.  These methods were programmed in R and the code is available at https://github.com/spirikahu.

\section{Case-control study}
\label{Section:case-control}

To illustrate our methodology we use the leptospirosis data given in Table~\ref{tab:lepto}, which was used to explore whether exposure to the bacterium \textit{leptospira} resulted in flu-like symptoms in New Zealand abattoir workers; for details see \cite{Dreyfus2014}. This was actually a cross-sectional study with a fixed total sample size of $380$, and was analysed as such in \cite{Pirikahu2016}. To illustrate how results for such data vary according to study design, we assume here that the data were instead collected in a case-control manner where the number of diseased and disease free individuals sampled is fixed by design. This means that the prevalence of disease cannot be estimated from this data. There are three population parameters defining the PAR, but under this study design we can only estimate two: $P(E^+|D^+)$ and $P(E^+|D^-)$. A Bayesian approach can be taken where a prior distribution is selected for the prevalence of disease, incorporating our knowledge and uncertainty regarding this parameter. Alternatively, experts might find it easier to instead specify a prior for $P(E^+)$. We explore both situations here. 

\begin{table}
	\begin{center}
		\begin{tabular}{lllr}
			\toprule
			\multicolumn{1}{r}{} & \multicolumn{2}{c}{Diseased} \\
			Exposed &  $D^+$  & $D^-$ & Total \\
			\hline
			$E^+$     & $22$    & $25$ & $47$ \\
			$E^-$    & $82$    & $251$  & $333$    \\
			\hline
			Total & $104$   & $276$  & $380$\\
			\bottomrule
		\end{tabular}
	\end{center}
	\caption{Data from the New Zealand leptospirosis study for sheep abattoirs performed by \cite{Dreyfus2014}. $E^+$ indicates exposure to the bacterium {\it leptospira} and $D^+$ the presence of flu-like symptoms.}
	\label{tab:lepto}
\end{table}

To model our case-control data let $x_{ij}$ represent the observed counts in a $2 \times 2$ table where $i \in \{1,2\}$ represents the row number and $j \in \{1,2\}$ the column number of the table. As this is a case-control study $n_1 = x_{11} + x_{21}$ and $n_2 = x_{12} + x_{22}$ are fixed by design. If we let the random variable $X_{ij}$ represent the possible number of observations in the $i_{th}$ row and $j_{th}$ column of the table, then the appropriate statistical model is the product of the independent binomial distributions 
\begin{equation}\label{Model:case-control}
X_{11} \sim \mbox{Binomial}(n_1, \phi_{1}) \quad\mathrm{and}\quad X_{12} \sim \mbox{Binomial}(n_2, \phi_{2}),  
\end{equation}
where $\phi_{1} = P(E^+|D^+)$ and $\phi_{2} = P(E^+|D^-)$. To calculate the PAR (and PAF) we require estimates for the population $P(D+)$ or $P(E+)$. In the following sections we describe an approach where prior information is specified on $P(D+)$ or $P(E+)$. The former case is very straight forward but that later is different and requires some care.

\subsection{Specifying prior information on disease prevalence}
\label{sect_CC_D+}

Let $\phi_3=P(D^+)$, $\phi = (\phi_1, \phi_2, \phi_3)$ and assign the priors $\phi_1 \sim \mbox{Beta}(\alpha_1,\beta_1)$, $\phi_2 \sim \mbox{Beta}(\alpha_2,\beta_2)$ and $\phi_3 \sim \mbox{Beta}(\alpha_3,\beta_3)$. Typically we might use uniform priors for $\phi_1$ and $\phi_2$ ($\alpha_1=\alpha_2=\beta_1=\beta_2=1$), but for $\phi_3$ we choose $\alpha_3 \ll \beta_3$ making the assumption that the disease is rare. Given the underlying model (\ref{Model:case-control}) and the fact that beta and binomial distributions are conjugate, the joint posterior distribution for $\phi$, calculated by multiplying the likelihood for the model by the priors, simplifies to:
\begin{equation}\label{eqn:casecontrolpost}
p(\boldsymbol{\phi}|n_1, n_2, X) \propto \phi_3^{\alpha_3 -1} (1-\phi_3)^{\beta_3 -1} \prod_{i=1}^2 \phi_i^{x_{1i} + \alpha_i-1} (1-\phi_i)^{\beta_i + n_i - x_{1i}-1}. 
\end{equation}
The joint posterior distribution (\ref{eqn:casecontrolpost}) is a product of independent beta distributions for each component of $\phi$. The prior for $\theta_3$ simply becomes the posterior, so no information can be gained about $\phi_3$ from the data. The distributions for $\phi_1$ and $\phi_2$ are: 
\begin{align}\label{eqn:post_phi12}
p(\phi_1|n_1,X) &\sim \mbox{Beta}(\alpha_1 + x_{11}, \beta_1 + n_1 - x_{11}), \quad \mbox{and}\nonumber\\
 p(\phi_2|n_2,X) &\sim \mbox{Beta}(\alpha_2 + x_{12}, \beta_2 + n_2 - x_{12}).
\end{align}
To calculate the PAR we require estimates of $p = P(D^+|E^+)$, $q = P(D^+|E^-)$ and $e=P(E^+)$. We can specify these parameters in terms of $\phi$ as follows:
\begin{align}\label{eqn:case-control p,q,e}
p & = \frac{\phi_1 \phi_3}{\phi_1 \phi_3 + \phi_2(1-\phi_3)}\\
q & = \frac{(1-\phi_1)\phi_3}{(1-\phi_1)\phi_3 + (1-\phi_2)(1-\phi_3)}\\
e & = \phi_1 \phi_3 + \phi_2(1-\phi_3).
\end{align}
The $PAR$ is then given by:
\begin{align}
PAR & = e(p-q) = \phi_1 \phi_3 - \frac{(1-\phi_1)\phi_3[\phi_1\phi_3 + \phi_2(1-\phi_3)]}{(1-\phi_1)\phi_3 + (1-\phi_2)(1-\phi_3)}.\label{eqn:case-control PAR}
\end{align}
The $PAF$ can also be calculated by simply dividing PAR by $\phi_3$. Re-sampling $\phi$ from its posterior distribution as given above (e.g.\ with the \texttt{rbeta} function in R), then allows samples from the posterior distributions of PAR and PAF to be obtained and summarized. 

In our leptosperosis example we used $\alpha_1 = \alpha_2 = \beta_1 = \beta_2 = \alpha_3 = 1$ and $\beta_3 = 1000$, which combined with the data (Table~\ref{tab:lepto}) gave posterior means and credible intervals, based on 10,000 iterations, of $PAR = 0.0013$ ($95\%$ CI: $0.00003, 0.005$) and $PAF = 0.14$  $(0.05, 0.23)$. The $PAR$ here represents the reduction in the risk to abattoir workers of experiencing flu-like symptoms that could be achieved by eliminating exposure to {\it leptospira}. The PAR is very small in this example because the assumed prevalence of disease is very low. For comparison the PAR for the cross-sectional study \citep{Pirikahu2016} was $0.04$ ($95\%$ CI: $0.009, 0.05$).

\subsection{Specifying prior information on the exposure rate}
\label{sec:CCE+}
In order to estimate $p$, $q$ and $PAR$ using equations~(\ref{eqn:case-control p,q,e}-\ref{eqn:case-control PAR}), $\phi_3$ must first be expressed in terms of $\phi_1, \phi_2$ and $e$ as follows:
\begin{equation*}\label{eqn:phi3_PD}
\phi_3 = \frac{e - \phi_2}{\phi_1 - \phi_2}.
\end{equation*}
Since $\phi_3$ represents a probability, it must be constrained to the interval $[0,1]$. This introduces the constraint that either $\phi_2 < e \leq \phi_1$ or $\phi_1 < e \leq \phi_2$. To account for these constraints let $A_{\phi} =[\min(\phi_1,\phi_2), \max(\phi_1,\phi_2)]$. When assigning priors for $\phi_1$, $\phi_2$ and $e$ one alternative is to specify each prior independently, then constraining these such that $e \in A_{\phi}$. For our particular example we assign $\mbox{Beta}(1,1)$ priors on $\phi_1$ and $\phi_2$ as before, and suppose there is prior information about the exposure rate specifying low exposure, $e \sim \mbox{Beta}(1,10)$. The joint posterior distribution for $\phi_1$, $\phi_2$ and $e$ can then be represented by:
\begin{equation*}   
p(\phi_1, \phi_2, e|n_1, n_2, X) \propto 
\begin{cases} 
	\begin{aligned}
		e^{\alpha_4 -1} (1-e)^{\beta_4 -1} \prod_{i=1}^2 \phi_i^{(x_{1i} + \alpha_i)-1} \\
			\times (1-\phi_i)^{(\beta_i + n_i - x_{1i})-1}, 
	\end{aligned} & \text{if $e \in A_{\phi}$}\\
0,  & \text{if $e \not\in A_{\phi}$}
\end{cases}
\end{equation*}
where in this example $\alpha_4 = 1$ and $\beta_4 = 10$. The full conditional distributions can be identified so we can adopt Gibbs sampling. For example the full conditional posterior distribution for $e|\phi$ is the $\mbox{Beta}(\alpha_4, \beta_4)$ distribution, truncated to $[\phi_1, \phi_2]$ when $\phi_1 < \phi_2$ or $[\phi_2, \phi_1]$ when $\phi_2 < \phi_1$. Sampling from this truncated beta distribution can be carried out by taking draws from $\mbox{Beta}(\alpha_4, \beta_4)$ then rejecting those values which do not fall inside the appropriate interval, or more efficiently by direct ``inverse-cdf'' sampling from the truncated $\mbox{Beta}(\alpha_4, \beta_4)$ distribution as described below. 

Sampling $\phi_1$ and $\phi_2$ is more complex as the parameter space is split into two regions: $\phi_1 > \phi_2$ and $\phi_2 > \phi_1$. To avoid the sampler being trapped in one of these regions, we propose joint sampling of these distributions. The unconstrained full conditionals are independent Beta distributions as in (\ref{eqn:post_phi12}); we sample from these until $\phi_1$, $\phi_2$ are on opposite sides of $e$, satisfying the constraint.

A comprehensive outline of the sampling procedure is given below. We denote the (unconstrained) prior distribution function for $e$ by $F_e(.)$ and its inverse by $F^{-1}_e(.)$. In \texttt{R} these are \texttt{pbeta()} and \texttt{qbeta()}.\\
\\
	1.  Specify initial values $\phi_1^0$ and $\phi_2^0$ such that $\phi_1^0 \neq \phi_2^0$, according to (\ref{eqn:post_phi12}), and initialise an iteration counter at $t=1$. \\
	2.  Calculate $A^t_L = \min(\phi_1^{t-1}, \phi_2^{t-1})$ and $A^t_U = \max(\phi_1^{t-1}, \phi_2^{t-1})$.\\
	3. Draw $u$ from $\mbox{Uniform}(F_e(A^t_L),F_e(A^t_U))$, then set $e^t = F^{-1}_e(u)$.\\
	4. Repeatedly draw $(\phi_1^{\dagger},\phi_2^{\dagger})$ from (\ref{eqn:post_phi12}) until $(\phi_1^{\dagger}-e^t)(\phi_2^{\dagger}-e^t)<0$, then set $\phi_1^{t}=\phi_1^{\dagger}$, $\phi_2^{t}=\phi_2^{\dagger}$.\\
	5. Set the iteration counter from $t$ to $t+1$.\\
	6. Repeat steps 2-5 until desired number of iterations is complete.\\

Performing 1,000 iterations, after a burn-in of 1,000, following the above procedure and applying the formulae (\ref{eqn:case-control p,q,e}-\ref{eqn:case-control PAR}) to estimate the PAR for the leptospirosis data, resulted in a mean estimate of $0.025$ ($95\%$ CI: $0.0018, 0.056$). The mean $PAF$ and its credible were also calculated; $PAF = 0.096$ ($95\%$ CI: $0.0074, 0.206$).	

\section{Cohort study}

Cohort studies involve following a group of individuals who share a similar characteristic, such as being exposed or not exposed to a certain risk factor, over a period of time. The numbers of individuals in the cohort who are exposed or not exposed to the risk factor of interest are fixed by design, meaning the probability of exposure cannot be estimated from the data. The methodology for estimating the PAR and its credible interval for a cohort study is very similar to that of the case-control study. We can either apply a prior distribution to the probability of exposure, $e$, or specify a prior for the prevalence of disease, $\phi_3$, which induces a prior distribution on $e$. We briefly explore both situations here but this time assume that the leptospirosis data (Table~\ref{tab:lepto}) was collected according to the cohort study design where $m_1 = x_{11} + x_{12}$ and $m_2 = x_{21} + x_{22}$ are fixed in advance. The appropriate statistical model in this case is the product of the following binomial distributions:
\begin{equation}\label{Model:cohort}
X_{11} \sim \mbox{Binomial}(m_1, p) \quad\mathrm{and}\quad X_{21} \sim \mbox{Binomial}(m_2, q).
\end{equation}

\subsection{Specifying prior information on the exposure rate}
Applying a prior to $e$ and deriving the posterior distribution for PAR can be done in a somewhat similar fashion to the case-control example where a prior was applied to the prevalence of disease, $\phi_3$ (Section~\ref{sect_CC_D+}). Let the priors on $e$, $p$ and $q$ be $\sim \mbox{Beta}(\alpha_4,\beta_4)$, $\mbox{Beta}(\alpha_5, \beta_5)$ and $\mbox{Beta}(\alpha_6, \beta_6)$ respectively. Given the underlying model (\ref{Model:cohort}) the joint posterior distribution for $p$, $q$ and $e$ calculated by multiplying the likelihood for the model and the priors is given by:
\begin{align*}
p(p,q,e|m_1, m_2, X) \propto \quad & e^{\alpha_4 -1} (1-e)^{\beta_4 -1} p^{x_{11}+\alpha_5-1} (1-p)^{\beta_5 + m_1 - x_{11} - 1}\\
& \times q^{x_{21} + \alpha_6 -1} (1-q)^{\beta_6 + m_2 - x_{21}-1}.
\end{align*}  
Note that the joint posterior is a product of independent beta distributions  where the marginal posterior distribution for $p$ and $q$ are:
\begin{align*}
p(p|m_1,X) &\sim \mbox{Beta}(\alpha_5 + x_{11}, \beta_5 + m_1 - x_{11}) \quad\mathrm{and} \\
\quad p(q|m_2,X) & \sim \mbox{Beta}(\alpha_6 + x_{21}, \beta_6 + m_2 - x_{21}).
\end{align*}
After taking random draws from the posterior for $p$, $q$ and $e$ the posterior for PAR can then be estimated using (\ref{eqn:PAR2}). The posterior for the PAF can also be generated by dividing the posterior draws for the PAR by $\phi_3 = pe + q(1-e)$. 

\subsection{Specifying prior information on the disease prevalence}
If we wish to specify a prior on $\phi_3$, say $\phi_3 \sim \mbox{Beta}(\alpha_3, \beta_3)$, then in order to estimate the PAR $e$ must first be expressed in terms of $p$, $q$ and $\phi_3$ as follows:
\begin{equation*}
e = \frac{\phi_3 - q}{p - q}
\end{equation*}
Since $e$ represents a probability it must be constrained to the interval $[0,1]$. This introduces the additional constraints that: $q \leq \phi_3 \leq p$ or $p \leq \phi_3 \leq q$. In practice it unlikely for $q > p$ unless $E^+$ represents a protective exposure such as vaccination. Similarly to the case-control example where a prior was placed on $e$ (see Section~\ref{sec:CCE+}) the constraints can be accounted for by truncating the joint posterior distribution and adopting a Gibbs sampling procedure to update parameters.  Joint sampling of these regions can be carried out analogously to that outlined for the case-control study in Section~\ref{sec:CCE+}, therefore we do not repeat it here.

\section{Cross-sectional study incorporating diagnostic testing}

In a cross-sectional study a random and representative sample, of size $n$, is taken from the population at a single point in time. When the data can be represented by a $2 \times 2$ table, as in Table~\ref{tab:lepto}, the appropriate model is $(X_{11}, X_{12}, X_{21}, X_{22}) \sim \mbox{Multinomial}(n, \pi)$, where $\pi$ is the vector of probabilities $(\pi_{11}, \pi_{12}, \pi_{21}, \pi_{22})$ corresponding to the cells in table. These probabilities can be expressed in terms of the population parameters as $\pi_{11}=pe$, $\pi_{12}=(1-p)e$, $\pi_{21}=q(1-e)$ and $\pi_{22} = (1-q)(1-e)$.
 
During collection of the leptospirosis data, each subject's exposure status to the \emph{leptospira} bacterium was determined via the imperfect microscopic agglutination test, so there is the possibility of a false positive or false negative results. To incorporate the uncertainty associated with this diagnostic test we need information on the test sensitivity ($Se$) and specificity ($Sp$). In this situation the observed data $x_{11}$, $x_{12}$, $x_{21}$ and $x_{22}$ represent the cross-classification of test status ($T$) and disease ($D$).

Let $\eta_{ij}$ for $i,j \in \{1,2\}$ represent the observed probabilities, i.e. $\eta_{11}=P(T^+ \cap\, D^+)$, $\eta_{12}=P(T^+ \cap\, D^-)$, $\eta_{21}=P(T^- \cap\, D^+)$ and $\eta_{22}=P(T^- \cap\, D^-)$, that can be estimated directly. These observed probabilities can be defined in terms of $Se$, $Sp$ and $\pi$ as follows:
\begin{align}
\eta_{11} & = Se\pi_{11} + (1-Sp) \pi_{21} && \eta_{12} = Se\pi_{12} + (1-Sp)\pi_{22}\nonumber\\
\eta_{21} & = Sp\pi_{21} + (1-Se) \pi_{11} && \eta_{22} = (1-Se)\pi_{12} + Sp \pi_{22}.
\label{Model:measure error pi}
\end{align}
Note that since $\eta_{22} = 1- \eta_{11} - \eta_{12} - \eta_{21}$, it is not independent of $\eta_{11}$, $\eta_{12}$ and $\eta_{21}$, so only three equations are actually needed. The model now becomes:
\begin{equation}\label{model:Multi eta}
(X_{11}, X_{12}, X_{21}, X_{22}) \sim \mbox{Multinomial}(n, \eta), 
\end{equation}
where $\eta$ is the vector $(\eta_{11}, \eta_{12}, \eta_{21}, \eta_{22})$. The data has 3 degrees of freedom, but we must estimate 5 different parameters ($p$, $q$, $e$, $Se$ and $Sp$) to calculate the PAR. The model is clearly non-identifiable so Frequentist methods will not work, but provided that good prior information is available for at least some of the parameters a Bayesian estimate can be obtained. The posterior distribution in this situation will not converge to a single point but rather a ridge in the parameter space, which in the limit of an infinite amount of data is known as the ``limiting posterior distribution'' \citep{Gustafson2005}. Geometrically, this can be thought of as the restriction of the prior to the maximum liklihood ridge in the parameter space. 

The priors we adopt for our example are:
\begin{align}\label{model:cross priors}
p,q \sim \mbox{Beta}(1,1) && e \sim \mbox{Beta}(2,2) \nonumber\\
Se \sim \mbox{Beta}(25,3) && Sp \sim \mbox{Beta}(30, 1.5).
\end{align} 
Those for $p$, $q$, $e$ correspond to a flat Dirichlet(1,1,1,1) on $(\pi_{11}, \pi_{12}, \pi_{21}, \pi_{22})$ and those for $Se$, $Sp$ reflect expert opinion for the diagnostic test used in the leptospirosis study. For this particular problem the joint posterior distribution can not be derived analytically, so we must resort to numerical approximation via simulation. Standard Markov chain Monte Carlo updaters (e.g. Gibbs or Metropolis-Hastings) can be very inefficient here with the problem worsening as the data size gets bigger and the slope of the posterior ridge becomes steeper \citep{Johnson2001}. 

\cite{Gustafson2015} provides a general importance-sampling approach for estimating the posterior distribution for non-identifiable models, which we adapt for this problem. However, this approach relies on being able to find a ``transparent re-parameterisation'' where the distribution for the data depends only on the identifiable parameters and not the non-identifiable parameters. Finding such a re-parameterisation can prove difficult, especially as the number of parameters increases so a MCMC approach that does not require re-parameterisation maybe preferable. We therefore propose as an alternative some novel MCMC samplers that take into consideration the shape of the posterior ridge to aid in the convergence of the Markov chain when the model is non-identified. We then compare these samplers in terms of effective sample size (ESS) and efficiency (i.e. ESS per second) with the more standard Gibbs, Metropolis-Hastings and Hamiltonian samplers.     

\subsection{Monte-Carlo importance sampler}
The idea behind importance sampling is to draw samples from a ``wrong'', but convenient, joint posterior distribution and then to correct for having choosen from the wrong distribution by multiplying by an appropriate weighting factor \citep{Kahn1955}.  When the model is non-identifiable Gustafson's approach requires an appropriate transparent re-parametrisation to be found, such that data depends only on the identifiable parameters $\Phi_I$, not non-identifiable parameters $\Phi_N$, and the joint prior density $p(\Phi_I, \Phi_N)$ can be evaluated \citep{Gustafson2015}. A convenient prior density, $p^*(\Phi_I, \Phi_N)$, can be selected by specifying a marginal density for $p^*(\Phi_I)$ that makes sampling easy, and then specifying the conditional density $p^*(\Phi_N|\Phi_I)$. A Monte Carlo sample of size $n$, denoted ($\Phi_I^i$, $\Phi_N^i$) for $i=1,\ldots,n$, can then be drawn from the posterior distribution arising from the convenience prior, $p^*(\Phi_I, \Phi_N) = p^*(\Phi_I) p^*(\Phi_N|\Phi_I)$.  Adjusting $p^*(\Phi_I, \Phi_N)$ by applying the weights: 
\begin{equation}\label{eqn:impw}  
w_i \propto \frac{p(\Phi_I^{i}, \Phi_N^{i})}{p^*(\Phi_I^{i}) p^*(\Phi_N^{i}| \Phi_I^{i})}
\end{equation}
scaled such that $\sum^N_{i=1} w_i = 1$, will represent the desired posterior distribution $p(\Phi_I, \Phi_N|x)$.

For our example the multinomial model in terms of the original parameter vector $\theta = (\pi_{11}, \pi_{12}, \pi_{21}, Se, Sp)$ can be described by the transparent re-parameterisation to $\Phi = (\eta_{11}, \eta_{12}, \eta_{21}, Se, Sp)$ according to (\ref{Model:measure error pi}) and modelled via (\ref{model:Multi eta}). The parameter $\Phi_I$ in this case is the observed probabilities $\eta$, which are obviously identifiable. To calculate the importance sampling weights, $w_i$, we propose adapting Gustafson's method. The problem with (\ref{eqn:impw}) is that the constraints on the conditional prior depend on the values of the identifiable part, so the normalizing constant in the conditional prior is a complex function of $\eta$. By replacing Gustafson's $p^*(\Phi_I^i)p^*(\Phi_N^i|\Phi_I^i)$ by an overall prior $p^*(\Phi)$ we can avoid this problem as the normalising constant is now fixed. Thus $w_i$ becomes  
\begin{equation*}
w_i \propto \frac{p(\Phi)}{p^*(\Phi)},
\end{equation*}
where $p(\Phi)=p(\theta)|\partial \theta/\partial \Phi|$ is the prior distribution induced on $\Phi$ by the actual prior on $\theta$ and $p^*(\Phi)$ the convenience prior specified on $\Phi$. The prior for $\Phi$ must be restricted to the set of values of $\Phi$, say $A$, for which $\pi_{ij} \in [0,1]$. We take the convenience prior for $\Phi$ as
\begin{equation}\label{eqn:impconvprior}
p^*(\Phi) \propto Se^{24}(1-Se)^2 Sp^{29}(1-Sp)^{0.5}\mathbbm{1}(A),
\end{equation}
where $\mathbbm{1}$ represents the indicator function which is $1$ when  $\pi_{ij} \in [0,1]$ and $0$ otherwise.  Note that the support is dependent on $\eta$.  By specifying the convenience prior on $\Phi$ by (\ref{eqn:impconvprior}), we find that the full posterior density is  
\begin{equation*}
p^*(\Phi | x) \propto \eta_{11}^{x_{11}} \eta_{12}^{x_{12}} \eta_{21}^{x_{21}} (1-\eta_{11}-\eta_{12}-\eta_{21})^{x_{21}} Se^{24}(1-Se)^2 Sp^{29}(1-Sp)^{0.5} \mathbbm{1}(A).
\end{equation*}
We can sample from this posterior by drawing $\eta$ from a Dirichlet$(x_{11}+1, \ldots, x_{22}+1)$, sampling $Se$ and $Sp$ from their beta priors (\ref{model:cross priors}), and rejecting any parameter sets that fail to satisfy the constraints $\pi_{ij} \in [0,1]$. The normalizing constant is now marginalized so can be ignored in the importance weights. The prior induced on $\Phi$ by the actual prior on $\theta$ is
\begin{equation*}\label{eqn:impactualprior}
p(\Phi) \propto Se^{24}(1-Se)^2 Sp^{29}(1-Sp)^{0.5} \mathbbm{1}(A) | \partial \theta / \partial \Phi |,
\end{equation*}
so the importance weights become $(Se + Sp -1)^{-2}$ when $\Phi \in A$ and zero otherwise. 

\subsection{Metropolis-Hastings, Gibbs and Hamiltonian samplers}
\textbf{Metropolis-Hastings (MH) sampler:} The MH algorithm begins with the selection of a proposal distribution, $Q()$. From this distribution we then propose a new candidate value $\theta^{\dagger}$ which is either accepted, with probability $\alpha$, as the next value $\theta^{t+1}$ in the chain, or rejected, with probability $(1-\alpha)$, with the current value $\theta^t$ retained as the next value. Here 
\begin{equation*}
\alpha = \min \left\{
 1,\frac{p(\theta^{\dagger})L(\theta^{\dagger})Q(\theta^t|\theta^{\dagger})} 
{p(\theta^t)L(\theta^t)Q(\theta^{\dagger}|\theta^t)} \right\}.
\end{equation*}
The choice of $Q()$ is often arbitrary. We adopt the commonly used random walk sampler where $Q(\theta^{\dagger}|\theta^t) \sim \mbox{Normal}(\theta^t, c \sigma^*)$ and $c$ is a scalar tuning parameter and $\sigma^*$ a fixed estimate of the posterior standard deviation. An appropriate value for $\sigma^*$ can be determined by calculating the standard deviation of the first 1,000 iterations of the chain, then selecting the tuning parameter $c$ from an arbitrary initial value to achieve an acceptance rate around 20-50\%  \citep{Christensen2010}.  The tuning parameters we used can be seen in Table~\ref{tab:tune}.

\textbf{Gibbs sampler:} The Gibbs sampler is of particular use for problems where the full conditional posterior distribution for a component $\theta_i$ given all other components $\theta$, i.e.\ $p(\theta_i |\theta_1^{t}, \ldots, \theta_{i-1}^{t}, \theta_{i+1}^{t-1}, \ldots, \theta_n^{t-1}, y)$ , can be sampled directly. For our particular example we follow a similar approach to \cite{Joseph1995} and introduce latent variables that represent the number of subjects correctly and incorrectly classified by the diagnostic test. A full outline of our Gibbs sampler and the full conditional posterior distributions can be found in the Appendix.

\textbf{Hamiltonian Monte Carlo algorithm (HMC):}  The HMC algorithm is an MCMC sampler which allows for more effective exploration of the parameter space by incorporating gradient information about the target distribution. The HMC algorithm is based on the Hamiltonian which in Physics is a function of a position vector $q$ and momentum vector $p$. In non-physical applications of HMC $q$ corresponds to the parameters of interest $\theta$, whereas $p$ represents artificially introduced auxiliary variables typically with independent Gaussian distributions. For the HMC algorithm the Hamiltonian function is expressed as $H(q,p) = U(q) + K(p)$ 
where $U(q)$ represents the ``potential energy'', which is taken to be minus the log posterior density of the distribution for $q$, and $K(p) = \sum_{i=1}^d p_i^2/2$ the ``kinetic energy''. The Hamiltonian equations 
\begin{align*}
\frac{d q_i}{dt} & = p_i \label{eqn:HM_q}\\
\frac{d p_i}{dt} & = - \frac{\partial U}{\partial q_i}. 
\end{align*}
are solved for $q$ numerically using the ``leapfrog'' method with the HMC algorithm; for a detailed outline of the algorithm see \cite{Neal2011}. An appropriate step size $\epsilon$, total number of leapfrog steps $L$ and gradient vector for the target distribution $\nabla U(q)$ must be specified. 

For our particular problem $\nabla U(q)$ is the $5 \times 1$ matrix of negated partial derivatives of the log posterior distribution of the model (\ref{model:Multi eta}) and priors (\ref{model:cross priors}) with respect to each parameter of $\theta$. The choice of $L$ and $\epsilon$ aims to balance the acceptance rate, compute time and exploration of the parameter space. A common approach is to simply perform some preliminary runs using different values for $L$ and $\epsilon$, then select $L$ and $\epsilon$ based on which run provides an acceptance rate between 20-50\% \citep{Neal2011}.

\subsection{New adapted random walk samplers}

The aim of these new MCMC samplers is to adapt the MH algorithm to encourage moves in the direction for which the maximum likelihood remains constant. This is achieved by specifying  the covariance matrix, $\Sigma^*$ for a multivariate normal proposal distribution based on the Jacobian matrix $J=\partial \eta/ \partial \theta$. The null singular vectors of $J$ are tangential to the likelihood ridge \citep{Jones2010}. To take larger steps in the directions for which the likelihood is changing most slowly, we could take $\Sigma^* \propto (J^{T} J)^{-1}$. However, for an non-identified model $J^{T} J$ is singular and therefore can not be inverted. To circumvent this problem we adopt the approach used in ridge regression \citep{Hoerl1970}, of adding a small positive quantity to the diagonal of the matrix $J^{T} J$. This small quantity has very little effect on the singular vectors and provides a matrix which can be inverted. Given this information we propose the following covariance matrix for the multivariate normal proposal distribution:
\begin{equation}\label{eqn:covar1}
\Sigma^* = c(\tau I + J^{T} J)^{-1},
\end{equation}    
where $c$ is a scaling constant, $I$ is the identity matrix and $\tau$ a small positive quantity added to achieve an invertible matrix. Note that this formulation does not take the amount of data into consideration.

An alternative is to use the standard asymptotic approximation to the covariance, the expected Fisher information, $I_E(\hat{\theta}) = E[-\partial^2 l(\theta)/\partial \theta \partial \theta^T] = J^T D J$, where $D = E[- \partial^2 l(\eta) / \partial \eta \partial \eta^T]$ is diagonal with elements $n^2 / x_{ij}$ as in \cite{Bishop1975}. Incorporation of the data in this way allows for the elements of the covariance matrix to adapt to the sample size which might make the sampler easier to tune. In addition we could also allow for the incorporation of prior information, so we also propose the alternative covariance matrix 
\begin{equation}\label{eqn:covar3}
\Sigma^* = c \left[\tau I + \left(J^T D J + \frac{\partial^2 \log p(\theta)}{\partial \theta \partial \theta^T} \right) \right]^{-1}.
\end{equation}   
In our example the priors are independent beta distributions so $\partial^2 \log p(\theta)/\partial \theta \partial \theta^T $ is diagonal with components $-[\alpha_i/\theta_i^2] - [\beta_i/(1-\theta_i)^2]$. Note that (\ref{eqn:covar3}) is equivalent to $\Sigma^* = c  [\tau I + \partial^2 \log p(\theta| x) / \partial \theta \partial \theta^T]^{-1}$, where $\log p(\theta|x)$ represents the log posterior distribution for $\theta$. 

A potential disadvantage to specifying the proposal distribution in this way is the increased computational burden, since we are required to re-calculate $\Sigma^*$ for each MCMC iteration. Therefore, even if the method explores the posterior more rapidly than other methods it may perform poorly in terms of efficiency. Additionally it requires specification of two tuning parameters; our choices of these can be seen in Table~\ref{tab:tune}.  

\subsection{Simulation study}
Each of the sampling methods described over the previous sections were applied to the leptospirosis data, where $n=380$, and for samples of size $n=3,800$ and $n=38,000$ (i.e. the leptospirosis data where each entry in Table~\ref{tab:lepto}
 is multiplied by 10 or 100 respectively), since the sample size can affect the convergence of the Markov chain for a non-identified model. BGR analysis was performed to assess the convergence of each method and a total of 100,000 iterations, including burn in, for each sampler was carried out. A tuning period was implemented pre-simulation for each MCMC method, for every sample size, so that no method would be disadvantaged by a poor choice of initial conditions. All methods were carried out in R \citep{R2020}, with the effective sample size (ESS) for MCMC methods calculated using the {\tt coda} package \citep{Coda}. The ESS for MCMC methods depends on the autocorrelation; $\mbox{ESS} = n/(1+2\sum_{k=1}^{\infty}\rho_k)$, where $n$ is the chain length and $\rho_k$ the lag $k$ autocorrelation \citep{Kass1998}. For importance sampling, it depends on the importance weights $w_i$ \citep{Kong1994} : $ESS = (\sum_{i=1}^n w_i)^2/\sum_{i=1}^n w_i^2$.

\begin{table}
\centering
\begin{tabular}{lll}
\toprule
$n=380$ & $\tau$ & $c$\\
\hline
MH-random walk & NA & 2.15 \\
MH-$\Sigma^* = c[\tau I + (J^{T} D J + p''(\theta))]^{-1}$ & 0.1 & 0.5\\
MH-$\Sigma^* = c[\tau I + J^{T} J]^{-1}$ & 0.2 & 0.00075\\
\bottomrule
$n=3800$ & & \\
\hline
MH-random walk & NA & 2.15 \\
MH-$\Sigma^* = c[\tau I + (J^{T} D J + p''(\theta))]^{-1}$ & 0.1 & 0.5\\
MH-$\Sigma^* = c[\tau I + J^{T} J]^{-1}$ & 0.1	& 0.00009\\
\bottomrule
$n=38000$ & & \\
\hline
MH-random walk & NA & 2.15\\
MH-$\Sigma^* = c[\tau I + (J^{T} D J + p''(\theta))]^{-1}$ & 0.1 & 0.3\\
MH-$\Sigma^* = c[\tau I + J^{T} J]^{-1}$ & 0.005 & 0.000005 \\
\bottomrule
\end{tabular}
\caption{Tuning parameters for MCMC approaches (excluded HMC), where MH-$\Sigma^* = c[\tau I + (J^{T} D J + p''(\theta))]^{-1}$ represents the adjusted random walk sampler with $\Sigma^*$ given by (\ref{eqn:covar3}), and MH-$\Sigma^* = c[\tau I + J^{T} J]^{-1}$ by (\ref{eqn:covar1}).
Note the random walk approach was implemented component-wise where the value for $c$ remained the same for each of the 5 parameters in $\theta$. Additionally, $D=I_E(\hat{\theta})$ in $\Sigma^*$ for the adapted MH-random walk approaches.}
\label{tab:tune}
\end{table}

\subsubsection{Simulation results}

The acceptance rates for each of the sampling methods performed on each sample size are given in Table~\ref{tab:Accept}. Acceptance rates less than 100\% for the importance sampling approach, suggests that approximately 12\% of the time a solution for $\pi_{ij}$ was outside $[0,1]$. Tuning the approaches with proposal variances involving $J$ presented difficulties, especially as the sample size increased. Specifically when $\Sigma^* = c[\tau I + J^{T} J]^{-1}$ both $\tau$ and $c$ need to be very small in order for the proposed $\theta^{\dagger}$ to be accepted at all. The HMC algorithm was also impossible to tuning for the sample sizes $n=3800$ and $n=38000$.

\begin{table}
\centering
\begin{tabular}{llllll}
\toprule
$n=380$ & $p$ & $q$ & $e$ & $Se$ & $Sp$\\
\hline
MC importance sampling &	87.2 & 87.2 & 87.2 & 87.2 &	87.2\\
MH-random walk & 43.1 & 45.2 & 30.5 & 42.3 & 30.2\\
HMC & 63.8 & 63.8 & 63.8 & 63.8 & 63.8\\
MH-$\Sigma^* = c[\tau I + (J^{T} D J + p''(\theta))]^{-1}$ & 28.2 & 28.2 & 28.2 & 28.2 & 28.2\\
MH-$\Sigma^* = c[\tau I + J^{T} J]^{-1}$ & 21.0 & 21.0 & 21.0 & 21.0 & 21.0\\
\bottomrule
$n=3800$ & & & & &\\
\hline
MC importance sampling & 87.5 & 87.5 & 87.5 & 87.5 & 87.5\\
MH-random walk & 16.8 & 42.9 & 15.2 & 33.7 & 15.2\\
HMC &&&&& \\
MH-$\Sigma^* = c[\tau I + (J^{T} D J + p''(\theta))]^{-1}$ & 23.7  & 23.7 & 23.7 & 23.7 & 23.7\\
MH-$\Sigma^* = c[\tau I + J^{T} J]^{-1}$  & 24.2	& 24.2 &  24.2 & 24.2 & 24.2\\
\bottomrule
$n=38000$ & & & & &\\
\hline
MC importance sampling & 87.4 & 87.4 & 87.4 & 87.4 & 87.4\\
MH-random walk & 32.9 & 40.4 &	27.9 & 26.6	 & 32.6\\
HMC &&&&&\\
MH-$\Sigma^* = c[\tau I + (J^{T} D J + p''(\theta))]^{-1}$ & 22.9 & 22.9 & 22.9 & 22.9 & 22.9\\
MH-$\Sigma^* = c[\tau I + J^{T} J]^{-1}$ & 28.2	& 28.2 & 28.2 & 28.2 & 28.2\\
\bottomrule
\end{tabular}
\caption{Percent acceptance rates for each method with a chain of length $100000$. Acceptance rates are removed from the table when the method did not converge within the $100000$ iterations for all parameters according to the BGR diagnostic, for the specified sample size, or when the method could not be tuned. For methods where block-wise updating has been adopted the acceptance rate will be the same for all parameters. The Gibbs sampler is not included here as the acceptance probability is $1$. For the adapted MH-random walk approaches $D = I_E(\hat{\theta})$ in $\Sigma^*$. }
\label{tab:Accept}
\end{table}

Table~\ref{tab:ESS} provides a comparison of the ESS per $1000$ iterations for each of the different samplers. What is overwhelmingly clear is that the importance sampling based method vastly out performs the MCMC methods. Even as the sample size becomes large (i.e.\ $n=38000$) the importance sampling approach provides a similar ESS to that seen when the sample size is $n=380$. In terms of computational efficiency the importance sampling approach is greatly superior, as can be seen in Table~\ref{tab:Efficency}. The downside of this approach in general however is the need for a transparent re-parameterization. In cases where such a re-parameterization is too difficult to determine (e.g.\ due to high dimensionality) and an MCMC approach adopted, then the choice of sampler should be based on the sample size.

\begin{figure}
\includegraphics[width=16cm, height=19cm]{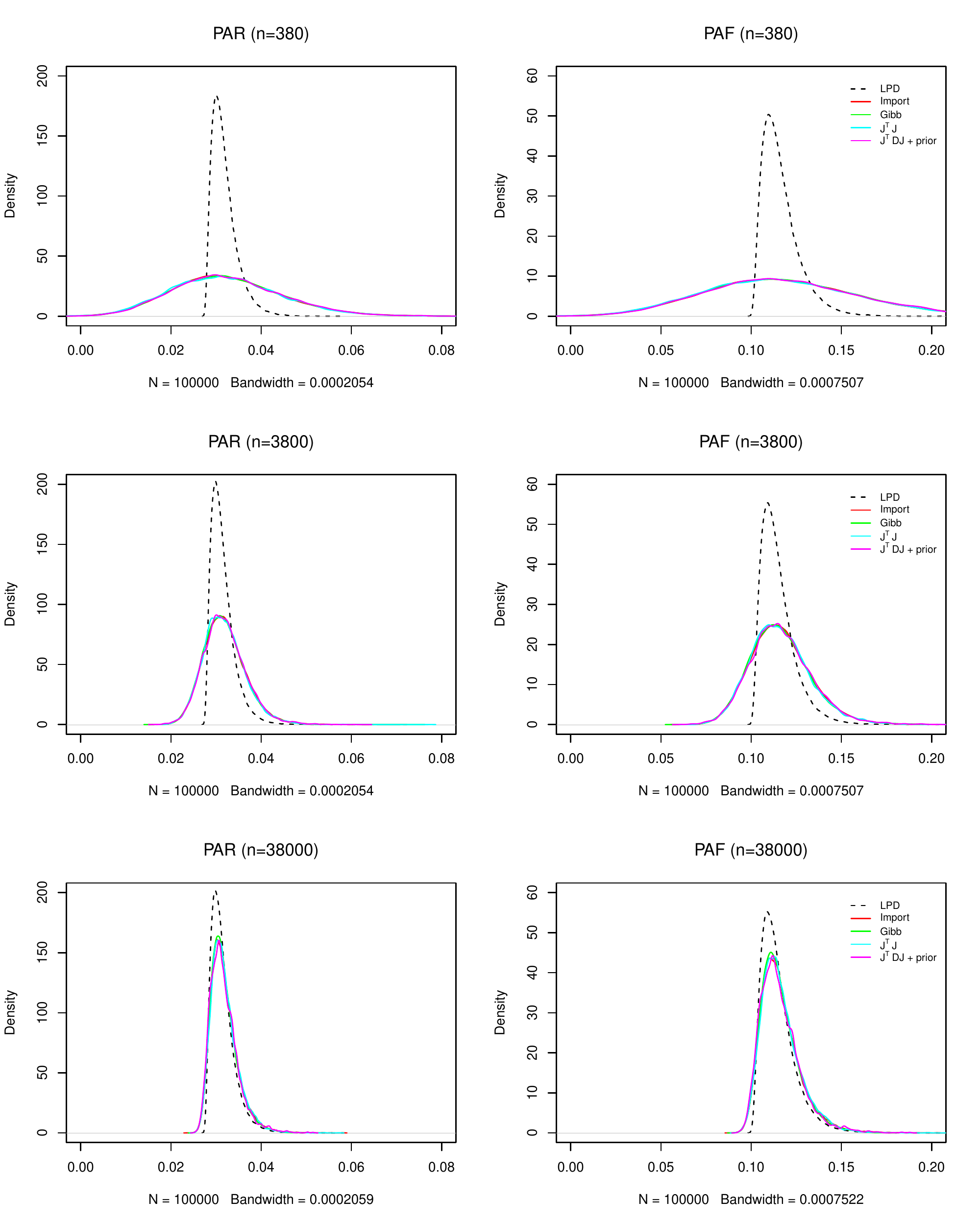}
\caption{Density plots for PAR and PAF comparing selected samplers with the LPD for differing sample sizes. Note that Import = Gustafson's importance sampler, Gibb = Gibbs sampler, $J^TJ$ = the adapted MH random walk sampler with $\Sigma^* = c[\sigma I + J^T J]$ and $J^T D J$ + prior = the adapted MH random walk sampler with $\Sigma^* = c[\tau I + (J^{T}I_E(\hat{\theta})J + p''(\theta))]^{-1}$ and LPD = limiting posterior distribution.}
\label{fig:posteriors}
\end{figure}

It can be seen that when $n=380$ the HMC sampler performs better than the other MCMC algorithms in terms of ESS for most parameters. However, Table~\ref{tab:Efficency} shows that this superior ESS comes at the cost of increased computational effort, in comparison to the random walk and Gibbs sampling approaches. The random walk and Gibbs sampler perform less well than HMC in terms of ESS when $n=380$ for most parameters, but better than the other MCMC methods investigated. Given their superior efficiency at $n=380$ and the ease with which they can be implemented, the random walk or Gibbs sampling approaches may be a viable option if a transparent parameterisation can not be found for implementation of Gustafson's approach or tuning an HMC algorithm presents difficulties. As the sample size increases the performance of the random walk and Gibbs sampler diminishes dramatically. This dramatic reduction in performance (in terms of ESS) for the random walk and Gibbs sampler occurs because as $n \rightarrow \infty$ the posterior ridge becomes narrower as it tends to the LPD. Figure~\ref{fig:posteriors} shows how the posterior distribution tends towards the LPD for the PAR and PAF, for selected samplers, as the sample size increases from $n=380$ to $n=38000$. The relatively wider posterior distribution we get when the sample size is small allows for larger steps in any direction to be taken without moving off the ridge. For the leptospirosis data ($n=380$) the estimate of the PAR under each of the methods was 0.03 ($95\%$ CI: 0.01-0.06) and the PAF 0.12 ($95\%$ CI: 0.04-0.21).

\begin{table}
\centering
\begin{tabular}{llllllll}
\toprule
$n=380$ & $p$ & $q$ & $e$ & $Se$ & $Sp$ & $PAR$ & $PAF$\\
\hline
MC importance sampling & 849.2 & 849.2 & 849.2 & 849.2 & 849.2 & 849.2 & 849.2\\
MH-random walk & 50.9 & 210.7 & 36.4 & 144.4 & 33.8 & 186.2 & 177.4\\
Gibbs sampler & 61.2 & 704.7 & 50.4 & 204.3 & 43.8 & 359.6 & 354.4\\
HMC & 225.3 &	465.2	& 127.9 & 335.2 & 133.0 & 144.6 & 163.3\\
MH-$\Sigma^* = c[\tau I + (J^{T} D J + p''(\theta))]^{-1}$& 28.2 & 29.3 & 26.7 & 29.7 & 26.4 & 30.4 & 29.8\\
MH-$\Sigma^* = c[\tau I + J^{T} J]^{-1}$ &  9.3 & 33.4 & 23.3 & 34.3  & 19.2 & 30.0 & 30.1\\
\bottomrule
$n=3800$ & & & & &&&\\
\hline
MC importance sampling & 851.7 & 851.7 & 851.7 & 851.7 & 851.7 & 851.7 & 851.7\\
MH-random walk & 2.7 & 106.4 & 3.0 & 34.7 &	2.7 & 46.5 & 46.6\\
Gibbs sampler & 4.9 & 154.6 & 5.1 & 27.8 & 4.6 & 57.5 & 57.1\\
HMC & &&&&&&\\
MH-$\Sigma^* = c[\tau I + (J^{T} D J + p''(\theta))]^{-1}$ & 8.7 &  24.5 & 10.0 & 22.6 & 9.5 & 24.2 & 24.1\\
MH-$\Sigma^* = c[\tau I + J^{T} J]^{-1}$  & 3.5 & 40.2 & 5.5 & 11.4 & 4.7 & 23.9 & 23.5\\
\bottomrule
$n=38000$ & & & & &&&\\
\hline
MC importance sampling & 851.3 & 851.3 & 851.3 & 851.3 & 851.3 & 851.3 & 851.3\\
MH-random walk & 0.5 & 7.7 & 0.6 & 5.2 & 0.6 & 5.2 & 5.2\\
Gibbs sampler & 0.5 & 5.5 & 0.6 & 3.2 &	0.5 & 3.6 & 3.6 \\
HMC &&&&&&&\\
MH-$\Sigma^* = c[\tau I + (J^{T} D J + p''(\theta))]^{-1}$ & 1.8 & 13.2 & 2.7 & 12.1 & 2.4 & 12.2 & 12.1\\
MH-$\Sigma^* = c[\tau I + J^{T} J]^{-1}$  & 3.6 & 23.3 & 5.5 & 16.0 & 5.1 & 18.7 & 18.4\\
\bottomrule
\end{tabular}
\caption{Effective sample size (ESS) per 1000 iterations. ESS values are removed from the table when the method does not converge within $100000$ iterations according to the BGR diagnostic, for the specified sample size, or when the method could not be tuned. Note that $D = I_E(\hat{\theta})$ in $\Sigma^*$ for the adapted MH-random walk approaches.}
\label{tab:ESS}
\end{table}

\begin{table}
\centering
\begin{tabular}{llllllll}
\toprule
$n=380$ & $p$ & $q$ & $e$ & $Se$ & $Sp$ & $PAR$ & $PAF$\\
\hline
MC importance sampling & 292.1 & 292.1 & 292.1 & 292.1 & 292.1 & 292.1 & 292.1\\
MH-random walk & 42.4 & 175.6 & 30.4 & 120.3 & 28.2 & 155.1 & 147.8\\
Gibbs sampler & 36.7 &	421.9 & 30.2 & 122.3 & 26.2 & 215.3 & 212.2\\
HMC & 41.2 & 85.0	& 23.4 & 61.3	& 24.3 & 26.4 & 29.8\\
MH-$\Sigma^* = c[\tau I + (J^{T} D J + p''(\theta))]^{-1}$ & 23.7 &	24.7 & 22.4 & 24.9 & 22.1 & 25.5 & 25.0 \\
MH-$\Sigma^* = c[\tau I + J^{T} J]^{-1}$ & 7.0 & 24.9 & 17.4 & 14.4 & 26.4 & 22.4 & 22.5\\
\bottomrule
$n=3800$ & & & & &&&\\
\hline
MC importance sampling & 313.0 & 313.0 & 313.0 & 313.0 & 313.0 & 313.0 & 313.0\\
MH-random walk & 2.4 &	96.7 &	2.8 & 31.6 & 2.5 & 42.3 & 42.4\\
Gibbs sampler & 1.5 & 48.9 & 1.6 & 8.8 & 1.5 & 18.2 & 18.1\\
HMC &&&&&&&\\
MH-$\Sigma^* = c[\tau I + (J^{T} D J + p''(\theta))]^{-1}$ & 8.5 & 24.0 & 9.8 & 22.2 & 9.3 & 23.7 & 23.6\\
MH-$\Sigma^* = c[\tau I + J^{T} J]^{-1}$ & 3.1 & 35.9 & 4.9 & 10.2 & 4.2 & 21.4 & 21.0\\
\bottomrule
$n=38000$&& & & & & &\\
\hline
MC importance sampling & 323.5 & 323.5 & 323.5 & 323.5 & 323.5 & 323.5 & 323.5\\
MH-random walk & 0.1 &	2.3 & 0.2 & 1.5 & 0.2 & 1.5 & 1.5\\
Gibbs sampler & 0.1 &	1.1 & 0.1 & 0.6 & 0.1 & 0.7 & 0.7\\
HMC &&&&&&&\\
MH-$\Sigma^* = c[\tau I + (J^{T} D J + p''(\theta))]^{-1}$  & 1.5 & 11.0 & 2.3 & 10.1 & 2.0 & 10.2 & 10.1\\
MH-$\Sigma^* = c[\tau I + J^{T} J]^{-1}$  & 2.7 & 17.4 & 4.1 & 11.9 &	3.8 & 14.0 & 13.7\\
\bottomrule
\end{tabular}
\caption{Effective samples performed per second (i.e.\ method efficiency). Efficiency values are removed from the table when the method does not converge within the $100000$ iterations according to the BGR diagnostic, for the specified sample size, or when the method could not be tuned. Note that $D = I_E(\hat{\theta})$ in $\Sigma^*$ for the adapted MH-random walk approaches.}
\label{tab:Efficency}
\end{table}

For the largest sample size, $n=38000$, the adapted random walk approach with proposal covariance matrix $\Sigma^* = c[\sigma I + J^TJ]$ is the preferred MCMC option,  performing slightly better in terms of ESS than all other MCMC based approaches. The elliptical shape of the proposal distribution appears to help the chain with exploring along the posterior ridge, although the low ESS suggests there is still a large amount of autocorrelation in the chain. The adapted random walk method with proposal covariance $\Sigma^* = c[\tau I + (J^{T}I_E(\hat{\theta}) J + p''(\theta))]^{-1}$ performed only slightly poorer, in terms of the ESS, than when $\Sigma^* = c[\sigma I + J^T J]$. The adapted  method with covariance $\Sigma^* = c[\tau I + (J^{T}I_E(\hat{\theta}) J + p''(\theta))]^{-1}$ but was simpler to tune due to adapting to the sample size. The slightly superior performance of the adapted approach with $\Sigma^* = c[\sigma I_n + J^T J]$ at $n=38000$ is likely a result of the smaller step size being taken. Computationally however, this approach can be quite intensive due to the matrix inversion required to provide the proposal variance, which is carried out for every iteration of the algorithm. For $n=38000$ though the effective samples generated per second for these method out-perform all other MCMC based methods.      

\section{Discussion}

The provision of confidence or credible intervals for PAR from a case-control or cohort study, allowing for all sources of uncertainty, has not been published previously to the best knowledge of the authors. Here we show that a Bayesian approach for estimating the PAR (and PAF) from a case-control or cohort study is very straightforward if beta priors are applied to the marginal probability of disease ($\phi_3$) or exposure ($e$) respectively. This is because the joint posterior distribution in these cases can be derived analytically. The obverse situations (case-control with prior on $e$; cohort study with prior on $\phi_3$) require a little more care because of constraints on the parameter space. We have proposed an MCMC sampler for these situations. The constraints also make the specification of priors difficult; rather than trying to elicit joint priors that respect the constraints, we propose a pragmatic approach in which ``independent priors'' are sought for each parameter without considering the constraints.

The cross-sectional leptospirosis study, where an imperfect diagnostic test was used to assess exposure status, gave rise to a much more complex example with a non-identified model. We have compared the performance of several different MCMC samplers, and developed a sampler which aims to effectively explore the posterior ridge of a non-identified model by taking into consideration the shape of the ridge. Comparison of effective sample size shows that the importance sampling approach proposed by \cite{Gustafson2015} was by far superior to all MCMC methods. It does however require a transparent parameterisation. If such a parameterisation is difficult to find or work with, MCMC simulation may be preferred. The choice of sampler in this situation should be based on the sample size of the data. When the sample size is small the HMC algorithm provided a greater number of effective samples per 1,000 iterations than the other MCMC samplers examined. Tuning the HMC algorithm though can be a difficult task, especially as the sample size increases. If the HMC algorithm cannot be tuned then the data-augmented Gibbs sampler provides the next best performance. For very large samples, our adapted random walk approach which takes into consideration the shape of the likelihood becomes competitive. Specifically, the adapted random walk approach with covariance matrix given by $\Sigma^* = c[\sigma I + J^T J]$ provides the greatest effective sample size. A strategy perhaps worth further investigation could be to alternate this sampler with the Metropolis-adjusted Langevin or HMC algorithms.

In general, analysts should be aware that standard MCMC updaters may not work well for Bayesian analysis of non-identified models, particularly for large datasets. This applies even for simple structures like the $2 \times 2$ table. Analogous results, and possible remedies, for more complex situations remain to be explored.

\subsection{ACKNOWLEDGEMENTS}
We are grateful to Cord Heuer for providing the data and priors, and to Matthew Schofield and Jonathan Marshall for suggestions that improved the efficiency of one of the algorithms.\\

\bibliography{UIM_SIMbib}

\begin{appendix}

\subsection{Gibbs sampler: Full conditional posterior distributions}
\label{appendix}
For our particular example (\ref{model:Multi eta}), Gibbs sampling requires the introduction of latent variables \citep{Joseph1995}. Let $Y_{ij}$ and $Z_{ij}$, where $i,j \in \{1,2\}$, be latent variables which represent the number of subjects that are correctly and incorrectly classified respectively. Additionally, it must hold that:
\begin{align}
& X_{11}=Y_{11}+Z_{21} && X_{12}=Y_{12}+Z_{22}\label{eqn: x11x12}\\
& X_{21}=Y_{21}+Z_{11} && X_{22}=Y_{22}+Z_{12}\label{eqn: x21x22},
\end{align}
recalling that $X_{ij}$ for $i,j \in \{1,2\}$ is what was actually observed. We can now express the likelihood for our model (\ref{model:Multi eta}) in terms of the latent variables $Y = (Y_{11}, Y_{12}, Y_{21}, Y_{22})$ and $Z = (Z_{11}, Z_{12}, Z_{21}, Z_{22})$ as:
\begin{multline}
L(X|Y, \pi, Se,Sp) \propto (\pi_{11}Se)^{Y_{11}} (\pi_{12}Se)^{Y_{12}} (\pi_{21}Sp)^{Y_{21}} (\pi_{22}Sp)^{Y_{22}} \times\\
 [(1-Sp)\pi_{12}]^{Z_{11}}\, [(1-Sp)\pi_{22}]^{Z_{12}} [(1-Se)\pi_{11}]^{Z_{21}}\, [(1-Se)\pi_{12}]^{Z_{22}},
\end{multline}
where $Z_{21} = X_{11} - Y_{11}$, $Z_{22}= X_{12} - Y_{12}$, $Z_{11} = X_{21} - Y_{21}$, $Z_{12}= X_{22} - Y_{22}$. Applying a $\mbox{Dirichlet}(1,1,1,1)$ prior on $\pi$, which is equivalent to applying the priors (\ref{model:cross priors}) specified for $p$, $q$ and $e$, the conditional posterior for $\pi$ is:
\begin{equation*}
p(\pi|X,Y, Se, Sp) \sim \mbox{Dirichlet}(Y_{11} + Z_{11} + 1, Y_{12} + Z_{12} + 1, Y_{21} + Z_{21} + 1, Y_{22} + Z_{22} + 1).
\end{equation*}
Given the priors (\ref{model:cross priors}) on $Se$ and $Sp$, the conditional posteriors for $Se$ and $Sp$ are:
\begin{align*}
p(Se| X, Y, \pi,Sp) \sim \mbox{Beta}(Y_{11} + Y_{12} + 25, Z_{11} + Z_{12} + 3)\nonumber\\
p(Sp|X, Y, \pi ,Se) \sim \mbox{Beta}(Y_{21} + Y_{22} + 30, Z_{21} + Z_{22} + 1.5).
\end{align*}
Finally the conditional posterior distributions for the latent variables $Y_{ij}$ are binomial:
\begin{align*}
P(Y_{11}|\pi, X,Se, Sp) &\sim \mbox{Binomial}\left(X_{11}, \frac{\pi_{11} Se}{\pi_{11}Se + (1-Sp)\pi_{21}}\right)\nonumber\\
P(Y_{12}|\pi, X,Se, Sp) &\sim \mbox{Binomial}\left(X_{12}, \frac{\pi_{12}Se}{\pi_{12} Se + (1-Sp)\pi_{22}}\right)\nonumber\\
P(Y_{21}|\pi, X,Se, Sp) &\sim \mbox{Binomial}\left(X_{21}, \frac{\pi_{21}Sp}{(1-Se)\pi_{12} + \pi_{21}Sp}\right)\nonumber\\
P(Y_{22}|\pi, X,Se, Sp) &\sim \mbox{Binomial}\left(X_{22}, \frac{\pi_{22} Sp}{(1-Se)\pi_{12} + \pi_{22}Sp}\right).
\end{align*}
The conditional posterior for the latent variables $Z_{ij}$ are also binomial, but in practice it is more efficient to determine $Z_{ij}$ using the relationships (\ref{eqn: x11x12}-\ref{eqn: x21x22}).  
\end{appendix}

\end{document}